\documentclass[aps,prd,twocolumn,superscriptaddress,eqsecnum,nofootinbib,showpacs,altaffilletter]{revtex4}
\usepackage{graphicx}
\usepackage{euscript,amssymb}
\usepackage{amsfonts}
\usepackage{amsmath}
\usepackage{amssymb}
\usepackage{graphicx}
\usepackage{euscript}
\usepackage{color}
\begin{document}

\title{Self-accelerating the normal DGP branch}

\author{Mariam Bouhmadi-L\'{o}pez}
\email{mariam.bouhmadi@ist.utl.pt}
\affiliation{Centro Multidisciplinar de Astrof\'{\i}sica - CENTRA, Departamento de F\'{\i}sica, Instituto Superior T\'ecnico, Av. Rovisco Pais 1,
1049-001 Lisboa, Portugal}


\begin{abstract}
We propose a generalised induced gravity brane-world model where the brane action contains an arbitrary $f(R)$ term, $R$ being the scalar curvature of the brane.
We show that the effect of the $f(R)$ term on the dynamics of a homogeneous and isotropic brane is twofold: (i) an evolving induced gravity parameter and (ii) a shift on the energy density of the brane. This new shift term, which  is absent on the Dvali, Gabadadze and Porrati (DGP) model, plays a crucial role to self-accelerate the generalised normal DGP branch of our model. We analyse as well the stability of de Sitter self-accelerating solutions under homogeneous perturbations and compare our results with the standard 4-dimensional one. Finally, we obtain power law solutions which either correspond to conventional acceleration or super-acceleration of the brane. In the latter case, no phantom matter is invoked on the brane nor in the bulk.  

\end{abstract}

\date{May 12, 2009}
\maketitle

\section{Introduction}

Understanding the recent acceleration of the universe is a challenging task facing the cosmological and particle physics community. The first evidence for the acceleration of the universe was provided  by the analysis  of the Hubble diagram
 of SNe Ia a decade ago \cite{Perlmutter:1998np}. This discovery, together with
 (i) the measurement of the fluctuations in the cosmic microwave background
 radiation (CMB) which implied that the universe is (quasi)  spatially flat
 and (ii) that the amount of matter which clusters gravitationally is much
 less than the critical energy density, implied the existence of
 a \textit{dark energy component} that drives the late-time acceleration
 of the universe. Subsequent precision measurements of the CMB
 anisotropy by WMAP \cite{Spergel:2003cb} and the power spectrum
 of galaxy clustering by the 2dFGRS and SDSS
 surveys \cite{Cole:2005sx,Tegmark:2003uf} have confirmed this discovery.

A possible approach to describing the late-time acceleration of the universe is to consider a modification of gravity, such that a weakening of this interaction on the appropriate scales induces the recent speed up of the universe (cf. Refs.~\cite{Nojiri:2006ri,Capozziello:2007ec,Sotiriou:2008rp,Durrer:2008in}). In other words, the weakening of gravity on large scales would provide an \textit{effective negative pressure} that would fuel the late-time acceleration of the universe.

The simplest of these models is the Dvali, Gabadadze and Porrati (DGP) scenario \cite{dgp,brane,reviewDGP}, which corresponds to a  five-dimensional (5D) model. In this model, our universe is a brane; i.e. a 4D hyper-surface, embedded in a Minkowski space-time. Matter is trapped on the brane and only gravity experiences the full bulk. The DGP model has two branches of solutions: the self-accelerating branch and the normal one. The self-accelerating brane, as its name suggests, speeds up at late-time without invoking any unknown dark energy component. On the other hand, the normal branch requires a dark energy component to accommodate the current observations \cite{normaldgp,Lazkoz:2006gp}. Despite the nice features of the self-accelerating DGP branch, it suffers from serious  theoretical problems like the ghost issue \cite{Koyama:2007za}. The main aim of this paper is to consider  a mechanism to self-accelerate the normal branch which is known to be free from the ghost issue \cite{Koyama:2007za}.
 
This mechanism will be based on a modified Hilbert-Einstein action on the brane and the simplest gravitational option is an $f(R)$ term. Extended theories of gravity based on 4D $f(R)$ scenarios have gathered a lot of attention in the last years (cf. the extensive lists of references in \cite{Nojiri:2006ri,Capozziello:2007ec,Sotiriou:2008rp}). It has been shown that these 4D models should follow closely the expansion of  a LCDM universe \cite{Hu:2007nk,Starobinsky:2007hu,Cognola:2007zu} and could have distinctive signatures on the large scale structure of the universe \cite{Song:2006ej,Pogosian:2007sw}. On the other hand, several methods have been invoked to reconstruct the shape of $f(R)$ from observations \cite{Capozziello1,Nojiri:2006gh,Capozziello2}, for example, by  using the dependence of the Hubble parameter with redshift which can be retrieved from astrophysical observations. We will show that an $f(R)$ term on the brane action can induce naturally self-acceleration on the normal DGP branch.

This paper is outlined as follows. In section \ref{sec2}, we present our model and deduce the modified Einstein equation on the brane. In section \ref{sec3}, we describe the dynamic of a homogeneous and isotropic brane on the framework introduced previously. Then we describe the effect of an $f(R)$ term on a standard induced gravity brane. On the next section \ref{sec4}, we obtain self-accelerating solutions corresponding to de Sitter space-times for both generalised DGP branches; i.e. the generalised normal solution (now self-accelerating) and the generalised self-accelerating solution. We study the stability of these solutions under homogeneous perturbations. We compare also our results with the standard 4D ones. In  section \ref{sec5}, we construct power law solutions for the brane expansion. Finally, in the last section we summarise and conclude.
\section{A Generalised induced gravity scenario}\label{sec2}

We consider a  brane, described by a 4D hyper-surface ($h$, metric g), embedded in a 5D bulk space-time ($\mathcal{B}$, metric 
$g^{(5)}$), whose action is given by 
\begin{eqnarray}
\mathcal{S} &=& \,\,\, \int_{\mathcal{B}} d^5X\, \sqrt{-g^{(5)}}\;
\left\{\frac{1}{2\kappa_5^2}R[g^{(5)}]\;+\;\mathcal{L}_5\right\}\nonumber \\
&& + \int_{h} d^4X\, \sqrt{-g}\; \left\{\frac{1}{\kappa_5^2} K\;+\;\mathcal{L}_4 \right\}\,, \label{action}
\end{eqnarray}
where $\kappa_5^2$ is the 5D gravitational constant,
$R[g^{(5)}]$ is the scalar curvature in the bulk and $K$ the extrinsic curvature of the brane in the higher dimensional
bulk, corresponding to the York-Gibbons-Hawking boundary term~\cite{York:1972sj}. For simplicity, we will assume that the bulk contains only a cosmological constant; i.e. $\mathcal{L}_5=-U$. Therefore, the bulk space-time geometry is described by an Einstein space-time 
\begin{equation}
G_{AB}[g^{(5)}]=-\kappa_5^2 U g_{AB}^{(5)}\,.
\label{Einsteinbulk}\end{equation}

The 4D Lagrangian corresponds to
\begin{equation}
\mathcal{L}_4=
\alpha f(R) + \mathcal{L}_m,
\label{branelagrangian}
\end{equation}
where $R$ is the scalar curvature of the induced metric on the
brane, $g$, and  $\alpha$ is a constant that measures the strength of the generalised induced gravity term $f(R)$ and has mass square units. Notice that therefore  the function $f(R)$ has mass square units.
 On the other hand, $\mathcal{L}_m$ corresponds to the matter Lagrangian of the brane which in particular may include a brane tension.  
The previous action, includes as a particular case the DGP model \cite{dgp,brane} when the bulk is flat, $f(R)=R$ and $\alpha=1/2\kappa_4^2$ where $\kappa_4^2$ is proportional to the 4D gravitational constant. For the time being we will consider
$f$ an arbitrary function of the scalar curvature of the brane.

As we are mainly interested in the cosmology of a homogeneous and isotropic brane it is quite useful to follow the approach introduced by Shiromizu, Maeda and Sasaki in\footnote{For an alternative approach to deduce the equations of evolution of  a DGP brane with curvature modifications on the brane action see \cite{Atazadeh:2007gs,Saavedra:2008qx}. See also \cite{Nojiri:2004bx} for a brane-world model with an $f(R)$ term.} 
 \cite{SMS}.  Then, the  projected Einstein equation
on the brane reads, where we have assumed a $\mathbb{Z}_2$-symmetry across the
brane, 
\begin{equation}
G_{\mu\nu}[g]=-\frac12 \kappa_5^2 U g_{\mu\nu} +\kappa_5^4
{{\Pi}}_{\mu\nu}-E_{\mu\nu}. \label{Einsteineq}
\end{equation}
Here, $\Pi_{\mu\nu}$ corresponds to the quadratic
energy momentum tensor  \cite{SMS}
\begin{eqnarray}
\Pi_{\mu\nu}=-\frac14
\tau_{\mu\sigma}{\tau_{\nu}}^{\sigma}+\frac{1}{12}\tau
\tau_{\mu\nu}+\frac{1}{8}
g_{\mu\nu}(\tau_{\rho\sigma}\tau^{\rho\sigma}-\frac{1}{3}\tau^2)
\,, \label{quadratic}\nonumber\\ \end{eqnarray}
and $E_{\mu\nu}$ is the (trace-free) projected Weyl tensor on the
brane.

The total energy momentum on the brane is defined as  
\begin{eqnarray}
\tau_{\mu\nu} \equiv -2\frac{\delta {\mathcal{L}}_4}{\delta
g^{\mu\nu}}+ g_{\mu\nu}{\mathcal{L}}_4,
\label{deftau}
\end{eqnarray}
and can be split into two terms 
\begin{equation}
\tau_{\mu\nu}=\tau_{\mu\nu}^{(m)} +\tau_{\mu\nu}^{(f)}.
\end{equation}
The first term  $\tau_{\mu\nu}^{(m)}$ corresponds to the energy momentum tensor of matter (which include in particular the brane tension) on the brane.  The second term
\begin{eqnarray} 
\tau_{\mu\nu}^{(f)}=
-&&2\alpha
\left\{\frac{df}{dR}R_{\mu\nu}-\frac12f(R)g_{\mu\nu}\right.\\
&&-\left.\left[{g_\mu}^\alpha
{g_\nu}^\beta-g_{\mu\nu}g^{\alpha\beta}\right]\nabla_\alpha\nabla_\beta\left(\frac{df}{dR}\right)\right\}.\nonumber
 \label{deftaug}
\end{eqnarray}
corresponds to the energy momentum tensor due to the generalised induced gravity term, $f(R)$,  on the brane. Now, if $f$ is proportional to the scalar curvature of the brane, then $\tau_{\mu\nu}^{(f)}$ is proportional to the Einstein tensor of the brane; i.e. the standard induced gravity brane-world scenario is retrieved:
\begin{equation}
\tau_{\mu\nu}^{(f)}=-2\alpha G_{\mu\nu}.
\end{equation}

Using the 5D Codacci equation, the bulk Einstein equation, and the junction condition at the brane, it turns out that  the total energy momentum tensor of the brane is conserved  $\tau_{\mu\nu}$ \cite{SMS}, i.e.
\begin{equation}
\nabla^\nu\tau_{\mu\nu}=0. \label{conservationtau}\end{equation}
On the other hand, because\footnote{We have proven this equation by making use of the 4D Bianchi identity on the brane; i.e. $\nabla^\nu G_{\mu\nu}=0$, and the relation between the non commutative character of two covariant derivatives and its relation to the Riemann curvature tensor (again on the brane), see for example equation 3.2.12 of Ref.~\cite{Wald}. Therefore, the conservation relation (\ref{conservationtauf}) can be proven in analogy to how it is done in the standard 4D $f(R)$ scenario.}
\begin{equation}
\nabla^\nu\tau_{\mu\nu}^{(f)}=0, \label{conservationtauf}
\end{equation}
we can then conclude that the energy momentum tensor of matter on the brane is conserved
\begin{equation}
\nabla^\nu\tau_{\mu\nu}^{(m)}=0. \label{conservationtaum}
\end{equation}

\section{dynamics of a FLRW brane}\label{sec3}

In what follows, we consider a Friedmann-Lema\^{\i}tre-Robertson-Walker (FLRW)  brane.
If the brane is homogeneous and isotropic it is known that the most general vacuum bulk in which the brane can be embedded is a Schwarzschild anti-de Sitter space-time \cite{Mukohyama:1999wi}. In this particular case, it is also known that the tensor $E_{\mu\nu}$ is conserved and by virtue of its traceless property, the projected Weyl tensor on the
brane behaves as a ``dark'' radiation fluid. 

The matter sector on the brane, prescribed by  $\tau_{\mu\nu}^{(m)}$, can be described by a perfect fluid with energy density $\rho^{(m)}$ and pressure $p^{(m)}$,  where $\rho^{(m)}$ is conserved as a consequence of Eq.~(\ref{conservationtauf}). On the other hand, an effective energy density and an effective pressure associated to $\tau_{\mu\nu}^{(f)}$ can be defined as follows
\begin{eqnarray}
\rho^{(f)}&=&-{\tau_0^0}^{(f)}\label{rhof}
\\&=&-2\alpha\left[3\left(H^2+\frac{k}{a^2}\right)f'-\frac12(Rf'-f)+3H\dot R f'' \right],\nonumber
\\
p^{(f)}&=&\phantom{-}{\tau_i^i}^{(f)} \label{pf}
\\&=& 2\alpha\left\{\left(2\dot H+3H^2+\frac{k}{a^2}\right)f'-\frac12(Rf'-f)\right.\nonumber\\
&& +\left.\left[\ddot R f''+(\dot R)^2f'''+2H\dot R f''\right]
\right\},\nonumber
\end{eqnarray}
where the label $i$ makes reference to the spatial coordinates of the brane. The parameter  $k=\pm 1, 0$ depending on the geometry  of the spatial sections of the brane. The dot refers to a derivative respect to the cosmic time of the brane $t$ and the prime refers to a derivative respect to the scalar curvature $R$. 

We recover the evolution equations of a  homogeneous and isotropic universe in the framework of 4D $f(R)$ theories  (given for example in the recent review \cite{Sotiriou:2008rp}) by simply imposing 
\begin{equation}
-\rho^{(f)}=\rho^{(m)};\quad -p^{(f)}=p^{(m)}.
\label{f(R)4d}
\end{equation}
This is a simple consequence of the definition (\ref{deftau}). In the standard 4D case the tensor (\ref{deftau}) has to vanish  due to the principle of least action because  $\mathcal{L}_4$  would be the total Lagrangian and therefore the variation of the 4D action with respect to the metric would have to be zero.

In the brane-world scenario things get a bit  more involved because of the  presence of (i) the quadratic energy momentum tensor $\Pi_{\mu\nu}$ and (ii) the projected Weyl tensor $E_{\mu\nu}$ in the modified Einstein equation on the brane (\ref{Einsteineq}). As we have already mentioned the effect of the projected Weyl tensor for a FLRW brane (embedded in a Schwarzschild anti-de Sitter space-time)  will be the presence of a ``dark''radiation fluid in the evolution  equations. On the other hand, the quadratic energy momentum tensor $\Pi_{\mu\nu}$ contributes to the 
effective Einstein equation (\ref{Einsteineq})
with 
\begin{eqnarray}
\Pi_0^0=-\frac{1}{12}\rho^2, \quad
\Pi_i^i=\frac{1}{12}\rho(\rho+2p).
\end{eqnarray}
In the previous equations
\begin{eqnarray}
\rho=\rho^{(m)}+\rho^{(f)}, \quad
p=p^{(m)}+p^{(f)}.
\label{rhop}
\end{eqnarray}
where $\rho^{(f)}$ and $p^{(f)}$ are given in Eqs.~(\ref{rhof}) and (\ref{pf}).
So, finally we can write down the modified Friedmann equation on the brane as
\begin{eqnarray}
\hspace*{-1cm}3\left(H^2+\frac{k}{a^2}\right)= \frac{\kappa_5^2U}{2}+ \frac{C}{a^4}+ \frac{\kappa_5^4}{12}\rho^2.\label{friedmann}
\end{eqnarray}
The second term on the right hand side (rhs) of equation (\ref{friedmann}) corresponds to the ``dark'' radiation energy density. The spatial component of Einstein equation can be expressed as       
\begin{eqnarray}
2\dot H+ 3H^2+\frac{k}{a^2}= \frac{\kappa_5^2U}{2}+\frac{C}{3a^4}
-\frac{\kappa_5^4}{12}\rho(\rho+2p),
\label{ray}
\end{eqnarray}
where the  energy density $\rho$ and the pressure $p$ are defined in Eq.~(\ref{rhop}). Even though Eqs.~(\ref{friedmann}) and (\ref{ray}) look very simple this is not case. For example, the brane Raychaudhuri equation (\ref{ray}) contains the effective quantities $\rho^{(f)}$ and $p^{(f)}$.

In order to compare this model with the standard induced gravity brane-world scenario  \cite{Maeda:2003ar}, i.e. $f(R)=R$ in the Lagrangian (\ref{branelagrangian}), it is useful to  define a rescaled induced gravity coupling
\begin{equation}
 \alpha^{(r)}=\alpha f'
\label{alphar}
\end{equation}
and split further the effective quantities $\rho^{(f)}$ and $p^{(f)}$ as
\begin{eqnarray}
\rho^{(f)}=\phantom{-}\rho^{\rm{(ig)}}+\rho^{(c)}, \quad p^{(f)}=p^{\rm{(ig)}}+p^{(c)},
\end{eqnarray}
where
\begin{eqnarray}
\rho^{\rm{(ig)}}&=&-6 \alpha^{(r)}\left(H^2+\frac{k}{a^2}\right), \\
p^{\rm{(ig)}}&=&\phantom{-}2\alpha^{(r)}\left(2\dot H+3H^2+\frac{k}{a^2}\right),
\end{eqnarray}
and
\begin{eqnarray}
\rho^{(c)}&=&2\alpha^{(r)}\left[\frac12\left(R-\frac{f}{f'}\right)-3H\dot{R}\frac{f''}{f'}\right], \label{rhoc}\\
p^{(c)}&=&2\alpha^{(r)}\left[\ddot R \frac{f''}{f'}+(\dot R)^2\frac{f'''}{f'}+2H\dot R \frac{f''}{f'}\right.\nonumber\\
&& \vspace*{0.2cm}\left. -\frac12\left(R-\frac{f}{f'}\right)\right].\nonumber\\
\end{eqnarray}

The magnitudes $\rho^{\rm{(ig)}}$ and $p^{\rm{(ig)}}$, we just defined, get  reduced to the effective energy density and pressure associated to the standard induced gravity models; i.e. models with $f(R)=R$. The effect of an $f(R)$ term on an induced gravity brane is  twofold: (i) an evolving induced gravity parameter, $\alpha^{(r)}$, and (ii) a shift on the matter energy density and matter pressure quantified by $\rho^{(c)}$ and $p^{(c)}$, respectively. In fact, we can rewrite the Friedmann equation as 
\begin{widetext}
\begin{eqnarray}
 \label{Hubble}
H^2+\frac{k}{a^2}=\frac{1}{6\alpha^{(r)}}\left\{\rho^{(m)}+\rho^{(c)}+
\frac{3}{\kappa_5^4\alpha^{(r)}}\left[1\pm\sqrt{1+\frac23\kappa_5^4\alpha^{(r)}\left(\rho^{(m)}+\rho^{(c)}-
\kappa_5^2\alpha^{(r)} U - 2\alpha^{(r)}\frac{C}{a^4}\right)}\right] \right\} \,.
\end{eqnarray}
The previous expression shows the existence of two branches of solution for $H^2$ as a
function of the effective energy density of the brane $\rho^{(m)}+\rho^{(c)}$.
On the other hand, the modified Raychaudhuri equation is
\begin{eqnarray}
\left\{1+ \frac{\kappa_5^4}{3}\alpha^{(r)}\left[\rho^{(m)}+\rho^{(c)}-6\alpha^{(r)}
\left(H^2+\frac{k}{a^2}\right)\right]\right\}\left(\dot
H-\frac{k}{a^2}\right)&=&-\frac{\kappa^4_5}{12}\left(\rho^{(m)}+\rho^{(c)}+p^{(m)}+p^{(c)}\right)\times  \label{Raychaudhuri} \\
&\times&
\left[\rho^{(m)}+\rho^{(c)}-6\alpha^{(r)}
\left(H^2+\frac{k}{a^2}\right)\right]-\frac23\frac{C}{a^4}\;.\nonumber
\end{eqnarray}
\end{widetext}
Thus the modified Einstein equations can be written similarly to that of a standard  induced gravity brane \cite{Maeda:2003ar}.  
The difference of course is hidden in $ \alpha^{(r)}$; i.e. an evolving induced gravity parameter, $\rho^{(c)}$ and $p^{(c)}$, which define an effective energy density and pressure on the brane:
\begin{equation}
\rho^{\rm{eff}}= \rho^{(m)}+\rho^{(c)}, \quad p^{\rm{eff}}= p^{(m)}+p^{(c)}.
\end{equation} 
A similar situation happens in an induced gravity scenario with an a non-minimally coupled scalar field on the brane \cite{BouhmadiLopez:2004ys}. 

A very important issue that we have not yet discussed is the relation between the effective 4D gravitational coupling and  the 5D gravitational constant $\kappa_5^2$. This issue  depends strongly on  which regime we are considering on the brane: early universe or late-universe (cf. for example \cite{brane,BouhmadiLopez:2004ys,BouhmadiLopez:2004ax}). The approach used in \cite{BouhmadiLopez:2004ys} can be extended to our model to retrieve the 4D effective gravitational constant at early-time on the brane.  Please, notice in that case one is considering a non vanishing brane tension. Here we are rather interested in the late-time evolution of the brane. Then, in this case, the brane effective gravitational constant can be easily obtained following the standard approach in the DGP model \cite{brane}. For simplicity, we choose  $k=0$, $U=0$ and  $C=0$, then the modified Friedmann equation (\ref{friedmann}) can be rewritten as
\begin{equation}
 H^2=\frac{1}{6\alpha^{(r)}} \left(\rho^{(m)}+\rho^ {(c)}\right) \pm \frac{1}{\kappa_5^2\alpha^{(r)}} H.
\label{DGPfriedmodified}
\end{equation}
Therefore, the effective gravitational constant on the brane is $8\pi G_{\rm{eff}}= 1/(2\alpha^{(r)})$ which is rescaled by an $f'(R)$ term with respect to the standard DGP model. On the other hand, we can define as well a new generalised crossover scale, $r_c^{(r)}$, which splits the 5D regime from the 4D regime\footnote{Here by a 5D regime we are referring to $H\propto\rho^{\rm{eff}}$; i.e. $r_c^{(r)} \ll H$, while in the 4D regime  $H^2\propto\rho^{\rm{eff}}$; i.e. $H\ll r_c^{(r)}$.}, such that  $r_c^{(r)}=\kappa_5^2\alpha^{(r)}$. 
Therefore, it is this quantity, $r_c^{(r)}$,  that  relates the gravitational constant on  the brane with  the fundamental 5D quantity $\kappa_5^2$. The precise shape and magnitude of $r_c^{(r)}$ has to be fixed by observation.  
Notice that $r_c^{(r)}$ is related to the crossover scale, $r_c$,  of the DGP model by $r_c^{(r)}=f'(R)r_c$.
On the other hand, the extra-dimension includes a new term ($\pm 1/(r_c^{(r)}) H$) in the modified Friedmann equation compared to 4D f(R) models (see Eq.~(\ref{DGPfriedmodified})).

Before concluding, we point out that the two branches of the DGP model \cite{dgp,brane} are particular cases of Eq.~(\ref{Hubble}) (and of course also of Eq.~(\ref{DGPfriedmodified})). These solutions correspond to $\alpha=1/(2\kappa_4^2)$, $f(R)=R$; therefore $\rho^{(c)}$ and $p^{(c)}$ would vanish, $C=0$ and $U=0$. These solutions in the absence of matter correspond to a de Sitter solution (self-accelerating branch) or a Minkowski space-time (normal branch). We will next show that the normal branch can become  self-accelerating due to the $f(R)$ contribution on the brane action.

\section{de Sitter  branes}\label{sec4}

One of the most puzzling problems nowadays in physics  is the issue of the late-time acceleration of the universe.
A possible approach to tackle this problem is within the frame-work of self-accelerating universes; i.e. could it be that a modification of gravity at late-time and on large scale be the cause of the current inflationary phase of the universe?
A de Sitter universe is  the simplest cosmological solution that exhibits acceleration and therefore it is worthwhile to prove the existence of this solution in our model and study its stability. This would be a first step towards describing in a realistic way the late-time acceleration of the universe in an $f(R)$ brane-world model. This approach will also enable us to look for self-accelerating solutions on the modified normal DGP branch ($-$ sign in Eq.~(\ref{Hubble})).

In this section, we first obtain the fixed points of the model corresponding to a de Sitter space-time and then we study their stability under homogeneous perturbations. For simplicity, we will consider the spatially flat chart of the brane; i.e. $k=0$, and no dark radiation on the brane; i.e. the bulk corresponds to a 5D maximally symmetric space-time. Notice that even in more general cases the dark radiation term will have no influence on the late-time dynamics of the brane as this term is constrained to be already subdominant by the time of nucleosynthesis \cite{Binetruy:1999hy}. 

\subsection{Background solutions}

\begin{figure}[h]
\begin{center}
\includegraphics[width=0.7\columnwidth]{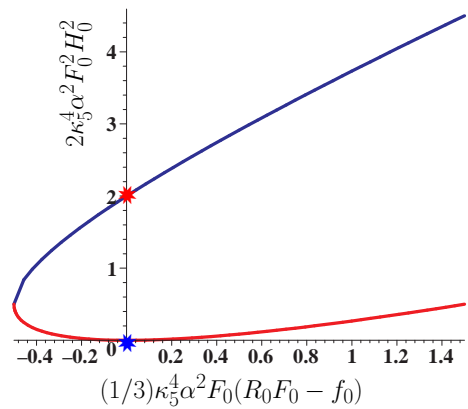}
\end{center}
\caption{The figure shows the behaviour of the rescaled squared Hubble rate $2\kappa_5^4\alpha^2F_0^2H_0^2$ for the two branches that generalise the DGP solution versus the rescaled energy density $\rho^{(c)}$ defined as $\frac13\kappa_5^4\alpha^2F_0(R_0F_0-f_0)$. The blue star corresponds to the normal DGP branch which is flat. The red star corresponds to the self-accelerating DGP branch. On the other hand, the blue curve corresponds to the generalised self-accelerating branch, while the red curve corresponds to the generalised normal branch.}
\label{ploth}
\end{figure}

\begin{figure}[h]
\begin{center}
\includegraphics[width=0.7\columnwidth]{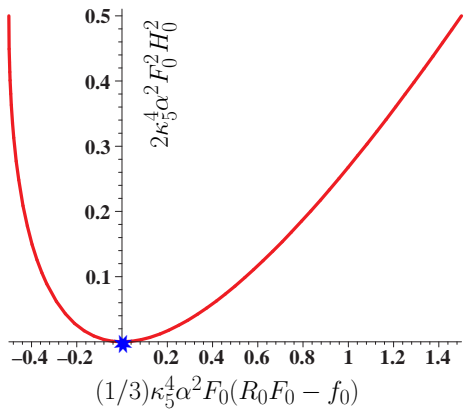}
\end{center}
\caption{This a zoom of the normal branch as it appears in figure \ref{ploth}.}
\label{ploth2}
\end{figure}

The Friedmann equation (\ref{friedmann}) implies that if the brane geometry corresponds to a de Sitter space-time with Hubble rate $H_0$, then
\begin{equation}
\rho^{(m)}-2\alpha\left(\frac12f -3H_0^2f'\right)=\rm{Constant}.
\end{equation}
In particular, the matter energy density has to be constant. Now imposing the conservation of the matter content of the brane, it turns out that matter in this case behaves like a cosmological constant $p^{(m)}=-\rho^{(m)}$. A constant matter energy density can always be re-absorbed in the $f(R)$ related terms. For this reason, we will disregard the matter content in our analysis of de Sitter branes. Finally, using Eq.~(\ref{Hubble}) we get that the Hubble parameter can be expressed as 
\begin{eqnarray}\label{desitterH1}
{2\kappa_5^4\alpha^2F_0^2}H_0^2&=&1 + \frac13\kappa_5^4\alpha^2F_0(R_0F_0-f_0)\nonumber\\
&&\hspace*{-0.8cm}+\epsilon\sqrt{1+\frac23\kappa_5^4\alpha F_0\big[\alpha(R_0F_0-f_0)-\kappa_5^2\alpha F_0U\big]}\nonumber\\
\end{eqnarray}
where $\epsilon=\pm 1$, the subscript 0 stands for quantities evaluated at the de Sitter space-time, $R_0=12 H_0^2$ and $F=df/dR$. 
We recover the DGP model for $f(R)=R$ and $U=0$. In fact, in that case, the de Sitter self-accelerating DGP branch is obtained for  $\epsilon= 1$ and the normal DGP branch or the  non-self-accelerating solution for $\epsilon=-1$. When the brane action contains curvature corrections to the Hilbert-Einstein action given by the brane scalar curvature, the branch with $\epsilon=-1$ is no longer flat and accelerates (cf. Figs \ref{ploth} and \ref{ploth2}). Therefore, an $f(R)$ term on the brane action induce in a natural way  self-acceleration on the normal branch. Most importantly, it  is known that such a branch is free from the ghost problem (see \cite{Koyama:2007za} and references therein). The reason behind the self-acceleration of the generalised normal brane is twofold: (i) $\rho^{(c)}$  defined in Eq.~(\ref{rhoc}) does not vanishes (in general) as it reduces to
\begin{equation}
\rho^{(c)}_0=\alpha(F_0R_0-f_0).
\end{equation}
and (ii)  $\rho^{(c)}$ enters the modified Friedmann equation (\ref{Hubble}) as an effective energy density.

Once we have seen the effect of an $f(R)$ term on the DGP model, we will ask the question the other way around: 
What is the effect of an extra-dimension on a 4D $f(R)$ model? In order to answer the previous question it is useful to rewrite the  Hubble rate (\ref{desitterH1}) as
\begin{equation}
H_0^2=H_{(4)}^2+\frac{1-\sqrt{1+\frac23\alpha^2\kappa_5^4F_0(f_0-\kappa_5^2UF_0)}}{2\alpha^2\kappa_5^4F_0^2},\label{desitterH2}
\end{equation}
where we have substituted $R_0=12H_0^2$ in Eq.~(\ref{desitterH1}) and  $H_{(4)}^2$ is defined as
\begin{equation}
H_{(4)}^2=\frac16\frac{f_0}{F_0}.
\end{equation}
The solution (\ref{desitterH2}) contains both DGP branches\footnote{Notice as well that the existence of two different branches is hidden on Eq.~(\ref{desitterH2}).}; i.e. 
solutions with $f(R)=R$, $U=0$ and $H_0^2=0, 1/(\kappa_5^4\alpha^2)$. 
The first term on the rhs of Eq.~(\ref{desitterH2}), $H_{(4)}^2$,  corresponds to the Hubble rate of de Sitter universes 
in 4D modified theories of gravity of the kind $f(R)$ (see for example \cite{Faraoni:2005vk,de Souza:2007fq}). Therefore, the presence of the extra dimension implies a shift on the Hubble rate (cf. Eq.~(\ref{desitterH2})). 

We notice as well that the de Sitter branes are close to the standard 4D regime as long as $H_0^2\sim H_{(4)}^2$, which implies
\begin{equation}
\left\vert\frac{3\left(1-\sqrt{1+\frac23\kappa_5^4\alpha^2F_0(f_0-\kappa_5^2UF_0)}\right)}{\kappa_5^4\alpha^2F_0f_0}\right\vert\ll 1.
\label{4dregime}
\end{equation}
This relation will be helpful for the next analysis.

\subsection{Stability analysis}

We next analyse the stability of de Sitter solutions under homogeneous perturbations up to first order on\footnote{We do not intend to analyse the ghost issue present in the self-accelerating DGP model in the present paper as it is beyond the scope of the present paper.}
\begin{equation}
\delta H= H(t)-H_0.
\end{equation}
We will follow the approach used in \cite{Faraoni:2005vk}. The perturbed Friedmann equation (\ref{friedmann}) implies an evolution equation for $\delta H$:
\begin{equation}
\delta \ddot H + 3H_0\delta \dot H + m_{\rm{eff}}^2\delta H=0,
\label{Friedmannpert}
\end{equation}
where the square of the effective mass $m_{\rm{eff}}^2$ is given by 
\begin{equation}
m_{\rm{eff}}^2= \frac13 \frac{F_0}{f_{RR}}-4H_0^2-\frac{1}{\kappa_5^4\alpha^2}\frac{1}{f_{RR}(f_0 -6H_0^2F_0)}.
\label{masseff}
\end{equation}
In the previous equation $f_{RR}={d^2f}/{dR^2}$, evaluated at the de Sitter background solution.

We restrict our analysis to $H_0>0$ as for $H_0<0$ there is always an exponentially growing mode for $\delta H$ implying  that the solution $H=H_0<0$ is unstable. Then, a de Sitter solution with $H_0>0$ is stable as long as $m_{\rm{eff}}^2$ is positive. 

It is quite useful to rewrite $m_{\rm{eff}}^2$ by substituting Eq.~(\ref{desitterH2}) into Eq.~({\ref{masseff}). We then get
\begin{equation} 
m_{\rm{eff}}^2= m_{(4)}^2+m_{\rm{shift}}^2+m_{\rm{pert}}^2,
\label{meff2}
\end{equation}
where
\begin{eqnarray}
 m_{(4)}^2&=&\frac13\left(\frac{F_0}{f_{RR}}-2\frac{f_0}{F_0}\right),\label{othermass}\\
m_{\rm{back}}^2&=&-\frac{2}{\alpha^2\kappa_5^4F_0^2}\left[1-\sqrt{1+\frac23\alpha^2\kappa_5^4F_0(f_0-\kappa_5^2UF_0)}\right],\nonumber \\
m_{\rm{pert}}^2&=&\frac{F_0}{3f_{RR}}\left[1-\sqrt{1+\frac23\alpha^2\kappa_5^4F_0(f_0-\kappa_5^2UF_0)}\right]^{-1}\nonumber.
\end{eqnarray}
The $m_{(4)}^2$ is the analogous quantity to $m_{\rm{eff}}^2$ in a 4D f(R) model \cite{Faraoni:2005vk}.  Therefore, the  extra-dimension induces a  shift on $m_{\rm{eff}}^2$ caused by two effects: (i) a purely background effect due to the shift on the Hubble parameter (see the second term on the rhs of Eqs.~(\ref{desitterH2}) and (\ref{masseff})) encoded on $m_{\rm{back}}^2$ and (ii) a purely perturbative extra-dimensional effect described by $m_{\rm{pert}}^2$.

In the remaining of this subsection, we will choose $F_0>0$, so the induced gravity parameter is positive (see Eq.~(\ref{alphar})). The last supposition guarantees a positive effective gravitational constant on the brane at late-time (cf. Eq.~(\ref{DGPfriedmodified})). Finally, we will also assume that we are slightly perturbing the Hilbert-Einstein action of the brane, i.e. $f_0\sim R_0$. Therefore, $f_0$ is positive because $R_0=12H_0^2$. For simplicity, we will assume as well  that $U=0$.
These three suppositions ($U=0$,  $f_0,\,F_0>0$) implies that the 4D regime (cf.~Eq.~(\ref{4dregime})) is reached when the inequality
\begin{equation}
1\ll\kappa_5^4\alpha^2F_0f_0
\label{4dregime2}
\end{equation}
is fulfilled. 

The previous inequality implies that  $m_{\rm{back}}^2>0$ and $m_{\rm{pert}}^2<0$. Consequently,  the shift on the brane Hubble parameter respect to the standard 4D case tends to make the perturbation heavier; i.e. the de Sitter universe would be more stable. However, the pertubative effect encoded on $m_{\rm{pert}}^2$ would make the perturbations lighter and therefore the de Sitter space-time would be less stable than in the pure 4D case. By imposing the condition (\ref{4dregime2}) on Eqs.~(\ref{meff2}) and (\ref{othermass}), it can be shown that the extra-dimension has  a \textit{benigner} effect in the 4D f(R) model; i.e. $m_{\rm{eff}}^2>m_{(4)}^2$, as long as 
\begin{equation}
{F_0^2}<{4f_0} f_{RR}.
\label{stabilitycondition}
\end{equation}


So far, we have  described the effect of the extra-dimension on the stability of a de Sitter solution under homogeneous perturbation in a 4D f(R) model. It is, however, not possible to analyse the effect of an f(R) term (on the brane action) on the stability of the de Sitter self-accelerating DGP solution. This is simply due to the fact that the f(R) contribution 
carries an extra degree of freedom which is absent in the original DGP model  or even GR models \cite{Hu:2007nk,Starobinsky:2007hu}. That is the reason why $m_{\rm{eff}}^2$ is not well defined for $f(R)=R$. The term $f_ {RR}$ vanishes in that case. This is not surprising as something similar happens in General Relativity (GR) and 4D f(R) models. Indeed,  $m_{(4)}^2$, the equivalent $m_{\rm{eff}}^2$ in the 4D f(R) scenario, is also not well defined for $f(R)=R$ in the standard 4D case which only means that the analysis does not apply to GR and not that de Sitter space-time is not stable in GR\footnote{We are grateful to V. Faraoni for pointing out this to us.}.

Finally, let us point out that the positiveness of  $m_{\rm{eff}}^2$ depends on which branch we are considering, as the Hubble rate $H_0$ depends on the specific branch [see Eq.~(\ref{desitterH1})]. So, a given self-accelerating solution  (\ref{desitterH1}) with $H_0>0$ is stable under homogeneous perturbations as long as  $m_{\rm{eff}}^2$ defined in Eq.~(\ref{masseff}) is positive.

\section{Power law expansion on the brane}\label{sec5}

In this section we construct power law solutions for the brane  expansion; i.e.  
\begin{equation}
a(t)=a_0\left(\frac{t}{t_0}\right)^{\beta},
\end{equation}
where $a_0, t_0, \beta\neq 0$ are constants\footnote{We henceforth disregard the case $\beta=0$ as it corresponds to a static brane which is not interesting from a cosmological point of view.} and $t$ is the cosmic time of the brane. For simplicity, we will consider that 
\begin{equation}
f=\tilde{f}_0 R^{n}, \,\,\,\tilde{f}_0, n=\rm{Constants}.
\end{equation}
These ans\"{a}tze have been previously used in standard 4D $f(R)$ models \cite{Capozziello:2002rd,Capozziello:2003gx}. 

Then the effective curvature energy density $\rho^{(f)}$ and pressure $p^{(f)}$, defined in Eqs.~(\ref{rhof}) and (\ref{pf}), satisfy
\begin{eqnarray}
{\rho^{(f)}}&=&6\left[\beta(n-2)+(2n-1)(n-1)\right]\frac{\alpha \beta f'}{nt^2},\\
{p^{(f)}}&=&2(2n-3\beta)\left[\beta(n-2)+(2n-1)(n-1)\right]\frac{{\alpha f'}}{nt^2},\nonumber \\
\end{eqnarray}
where 
\begin{equation}
 f'=n\tilde{f}_0\left[\frac{6\beta(2\beta-1)}{t^2}\right]^{n-1}.
\end{equation}
An equation of state parameter can be defined as
\begin{equation}
w_{f}=\frac{p^{(f)}}{\rho^{(f)}}=-1+\frac{2n}{3\beta}.
\label{wf}
\end{equation}
This parameter is not to be confused with the effective equation of state parameter $w_{\rm{eff}}$ defined through 
\begin{equation}
H^2=\frac{\kappa_{4}^2}{3}\rho_0 a^{-3(1+w_{\rm{eff}})},
\end{equation}
where $\rho_0$ is a constant. For a power law expansion $w_{\rm{eff}}$ reads 
\begin{equation}
w_{\rm{eff}}=-1+\frac{2}{3\beta}.
\label{weff}
\end{equation}

In standard 4D $f(R)$ models the quantities $\rho^{(f)}$ and $p^{(f)}$  vanish (see Eq.~(\ref{f(R)4d})) implying a set of constraints in the parameters $\beta$ and $n$  \cite{Capozziello:2002rd,Capozziello:2003gx}. In the brane-world scenario these constraints are different as a consequence of the modified Einstein equations (see Eqs.~(\ref{friedmann}), (\ref{ray})). Assuming that the brane is  spatially flat, i.e. $k=0$, and the bulk corresponds to a Minkowski space-time, i.e. $U=0$,
the Friedmann equation implies 
\begin{eqnarray} \label{friedmannpl}
\frac{\beta^2}{t^2}&=&\kappa_5^4\alpha^2\tilde{f}_0^2\left[\beta^2(n-2)+\beta(2n-1)(n-1)\right]^2\nonumber\\
&\times& \frac{\left[6\beta(2\beta-1)\right]^{2(n-1)}}{t^{4n}},
\end{eqnarray} 
while the Raychaudhuri equation imposes the constraint
\begin{eqnarray}\label{raychaudhuripl}
\frac{\beta(3\beta-2)}{t^2}&=&\kappa_5^4\alpha^2\tilde{f}_0^2\beta(3\beta-4n)\nonumber\\ &\times& 
\left[\beta(n-2)+(2n-1)(n-1)\right]^2\nonumber \\ &\times& 
\frac{\left[6\beta(2\beta-1)\right]^{2(n-1)}}{t^{4n}}.
\end{eqnarray}
From the previous two equations it turns out that 
\begin{eqnarray}
n&=&\frac12,\\
\beta&=&\frac{8}{16-3\kappa_5^2\alpha^2\tilde{f}_0^2}.
\end{eqnarray}
Then, we can conclude that $1+w_{\rm{eff}}=2(1+w_{f})$ (cf. Eqs.~(\ref{wf})-(\ref{weff})). The prefactor \textit{two} on the rhs of the previous equation is a brane effect; i.e. the energy density $\rho^{(f)}$ appears quadratically on the modified Friedmann equation.

The brane accelerates if  (i) $\beta>1$ or (ii) $\beta<0$. Case (i) corresponds to conventional acceleration and it will hold if 
\begin{equation}
1<\frac38\kappa_5^2\alpha^2\tilde{f}_0^2<2.
\end{equation}
Now, if the induced gravity parameter is defined roughly as in the DGP model \cite{dgp}; i.e. $\alpha\simeq 1/(2\kappa_4^2)=M_{P}^2$, we obtain bounds on the allowed set of $\tilde{f}_0$ values for the brane being accelerating
\begin{equation}
\frac{16}{3}\frac{M_5^3}{M_{P}^4}<\tilde{f}_0^2<\frac{32}{3}\frac{M_5^3}{M_{P}^4}.
\label{inequality1}
\end{equation}
This bound is related to the ratio of the fundamental Planck mass, $M_5$ , and the effective 4D Planck mass $M_{P}$.

Conversely, case (ii), which is given by 
\begin{equation}
 2<\frac38\kappa_5^2\alpha^2\tilde{f}_0^2,
\end{equation}
describes super-acceleration. By assuming that the induced gravity parameter is defined as in the DGP model \cite{dgp}, we get 
\begin{equation}
\frac{32}{3}\frac{M_5^3}{M_{P}^4}<\tilde{f}_0^2;
\label{inequality2}
\end{equation}
i.e. we obtain a  lower bound for $\tilde{f}_0$.

In summary, we have shown that the brane can follow a power law expansion such that the brane is accelerating ($1<\beta$) or even super-accelerating ($\beta<0$). In the latter case, the brane faces a big rip singularity in its future \cite{Caldwell:1999ew}. This happens at $t=0$ where the scale factor,  $H$, $\dot H$, $\ddot H$ and $\dddot H$ blow up. For $\beta<0$, we  have chosen the cosmic time to be negative so that the brane expands as time pass by. 
Unlike in a standard relativistic framework, in $f(R)$ models up to the third cosmic time derivative of the Hubble rate 
appear in the evolution equation of the universe. Consequently, it makes sense to speak about divergences of $\ddot H$  and its derivative. The super-acceleration of the brane is accompanied by a phantom-like behaviour; i.e. \mbox{$1+w_{eff}<0$}. Even the  effective fluid ($\rho^{(f)}$,$\,p^{(f)}$) mimics a phantom fluid; i.e. \mbox{$1+w_f<0$}, this  mimicry  is exclusively due to curvature effects and takes place although no real phantom matter has been considered in the model.

Alternative ways to get a phantom-like behaviour on the brane are based on a screening of the cosmological constant \cite{normaldgp,Lazkoz:2006gp,BouhmadiLopez:2008nf} or a flow of energy from the brane to the bulk \cite{{BouhmadiLopez:2005gk}}.  In the model presented here, the phantom mimicry is  a pure gravitational effect and it does not involve any of the two effects mentioned above.

\section{Conclusions}\label{sec6}

In this paper we present a mechanism to self-accelerate the normal DGP branch which unlike the original self-accelerating DGP branch is known to be free from the ghost problem\footnote{Notice as well that by embedding the DGP model in a higher dimensional space-time, the ghost issue present in the original DGP model may be cured \cite{deRham:2007xp} while preserving the existence of a self-accelerating solution \cite{Minamitsuji:2008fz}. See also Ref.~\cite{deRham:2006pe,Gabadadze:2006xm}}.  The mechanism is based in including  curvature modifications on the brane action. For simplicity, we choose those terms to correspond to an $f(R)$ contribution, which in addition is known to be the only higher order gravity theories that avoid the so called Ostrogradski instability in 4D models \cite{Sotiriou:2008rp}. 

We obtain the effective Einstein equation on the brane following the approach introduced  by Shiromizu, Maeda and Sasaki in Ref.~\cite{SMS}. We then describe the dynamic of a FLRW brane and  identify the two branches of solutions of the generalised induced gravity brane-world model. The $f(R)$ term on the brane action results on an evolving effective induced gravity parameter; i.e. a variable gravitational constant. The same term also induces a shift on the  energy density of the brane through the new magnitude $\rho^{(c)}$ (cf. Eqs.~(\ref{rhoc}) and (\ref{Hubble})). In the DGP model $\rho^{(c)}$ vanishes, however, for $f(R)\neq 0$ it is generally non zero. This is precisely the reason for getting self-acceleration in the branch that generalised the normal DGP solution (see Eq.~(\ref{desitterH1})).

We obtain all the  de Sitter self-accelerating solutions  and study their stability under homogeneous perturbations. It turns out that the self-accelerating solutions are stable as long as the parameter $m_{\rm{eff}}^2$, defined in Eq.~(\ref{masseff}) and related to the effective mass of the perturbation, is positive. This parameter can be splitted into three different terms: (i) the standard  4D contribution $m_{(4)}^2$ (see e.g. Ref.~\cite{Faraoni:2005vk}), a contribution from a purely background origin $m_{\rm{back}}^2$  and another one of perturbative origin $m_{\rm{pert}}^2$,  all defined in Eqs.~(\ref{meff2}) and (\ref{othermass}). For those solutions close to the standard 4D regime, i.e. those satisfying the inequality (\ref{4dregime2}), $m_{\rm{pert}}^2$ tends to make the homogeneous perturbation heavier ($m_{\rm{pert}}^2>0$), while $m_{\rm{back}}^2$ tends to make the same perturbation lighter ($m_{\rm{back}}^2<0$). Moreover, we have shown  that the extra-dimension has  a \textit{benigner} effect in a 4D f(R) model;  i.e. $m_{\rm{eff}}^2>m_{(4)}^2$, as long as  the inequality (\ref{stabilitycondition}) holds.

On the other hand, we have obtained power law solutions for the brane expansion which corresponds to conventional acceleration or super-acceleration. The super-acceleration is achieved without invoking any phantom matter on the brane or the bulk.

Last but not least, it is know that 4D $f(R)$ models are not free from theoretical problems (cf. Ref.~\cite{Straumann:2008ru} for a recent account on the subject), so in constructing an $f(R)$ brane-world model, we should of course try to avoid these theoretical troubles. We have just undertaken a first step towards constructing realistic self-accelerating solutions in the normal DGP branch. There are still many issues to be addressed, for example which $f(R)$ should  we pick up to be in agreement with the cosmological observations and the solar system tests? We leave these interesting issues for future works.

\acknowledgments

The author is grateful to Ruth Lazkoz for collaboration on the early stage of the paper. She also wishes to acknowledge the hospitality of  the Theoretical Physics group of the University of the Basque Country during the completion of part of this work.  The author is as well grateful to Salvatore Capozziello for very useful comments on a previous version of the paper. M.B.L. is  supported by the Portuguese Agency Funda\c{c}\~{a}o para a Ci\^{e}ncia e Tecnologia through the fellowship SFRH/BPD/26542/2006.

\end{document}